\documentclass[a4paper,11pt]{article}
\usepackage{pos}
\usepackage{wrapfig} 

\title{Automatic test system for the data acquisition board of GRANDProto300}

\author*[a,b]{Yiren Chen}
\author[a]{Jianhua Guo}
\author[a]{Shen Wang}
\onbehalf {for the GRAND collaboration}

\affiliation[a]{Purple Mountain Observatory Key Laboratory of Dark Matter and Space Astronomy, Chinese Academy of Sciences,\\
 10 Yuanhua Road , Nanjing, China}

\affiliation[b]{School of Astronomy and Space Science, University of Science and Technology of China,\\
96 jinzhai Road, Hefei, China}

\emailAdd{chenyiren@pmo.ac.cn}

\abstract{The Giant Radio Array for Neutrino Detection (GRAND) aims to detect the radio emission from air showers triggered by ultra-high-energy particles in the atmosphere. GRANDProto300 is its pathfinder array, of which the first 100 detection units have already been produced. A custom data acquisition (DAQ) board receives, processes, and transmits radio signals. We report on the board design, and on the functions and performance tests applied to it. Function tests are of signal acquisition and transmission, and of the field-programmable gate array (FPGA) filter algorithm. Performance tests are of the analog-to-digital conversion and GPS time accuracy. We developed an efficient system to automate the test, in line with the mass-production scale needed to build future, larger versions of GRAND.}

\ConferenceLogo{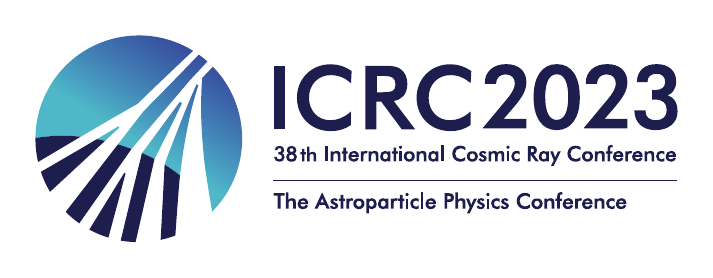}

\FullConference{%
38th International Cosmic Ray Conference (ICRC2023)\\
  26 July - 3 August, 2023\\
  Nagoya, Japan}

\begin{document}
\maketitle

\section{Detection concept}

The origin of ultra-high-energy cosmic rays (UHECRs) remains unknown since their discovery over fifty years ago. They are likely made in the most energetic cosmic accelerators \cite{ref1}. The discovery of the sources of UHECRs will bring new insight into the high-energy Universe \cite{ref2}. The direct strategy to search for their sources is to look for localized excesses in the arrival direction of UHECRs,  which is challenging because they are attenuated by cosmic photon backgrounds and deflected by intervening magnetic fields. The indirect strategy is to detect neutrinos and gamma rays made by UHECRs, which point directly back to their sources.

\begin{figure}[htbp]
\centering
\includegraphics[width=0.7\linewidth]{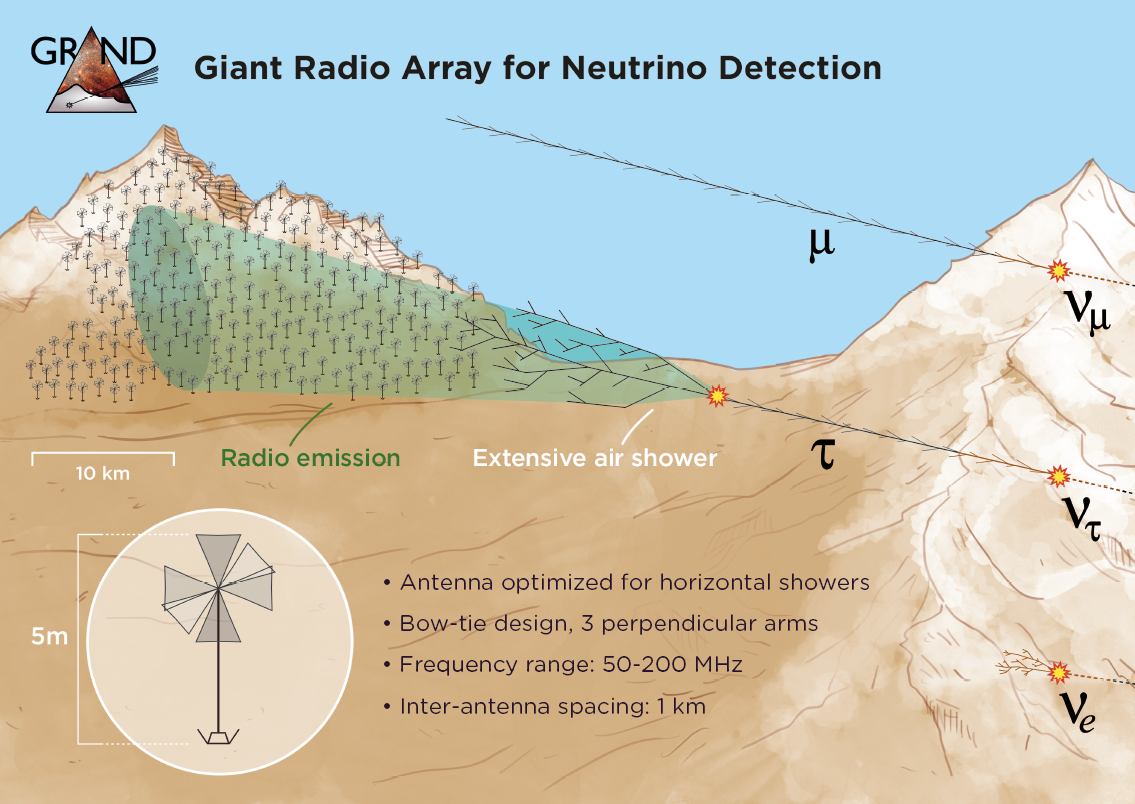}       
\caption{GRAND detection method \cite{ref5}.\label{Fig.1}}
\end{figure}

The envisioned Giant Radio Array for Neutrino Detection (GRAND) targets the detection of ultra-high-energy (UHE) cosmic rays, gamma rays, and especially, neutrinos, with energies above $10^{17}$~eV. The detection method is shown in Fig. \ref{Fig.1}. The antenna array is designed to trigger autonomously on the radio emissions from inclined extensive air-showers (EAS) made by UHECRs, gamma rays, and neutrinos. UHE tau neutrinos produce radio emissions by interacting with rocks, generating tau leptons that typically traverse tens of kilometers of rock before entering the atmosphere to decay and produce Earth-skimming EAS \cite{ref3,ref4}.

 The project will be modular – consisting of cheap, stable, and energy-efficient detection units and built in three stages. Starting from 2024, GRANDProto300 (GP300), a pathfinder array of 300 antennas, will demonstrate autonomously detection of inclined EAS and concentrate on the study of UHECRs with $10^{16.5}$--$10^{18.5}$~eV. After 2028, an array of 10,000 antennas, GRAND10K, will reach a neutrino sensitivity of 8×$10^{-9}$ GeV cm$^{-2}$ s$^{-1}$ sr$^{-1}$ as the sub-array of GRAND. In the 2030s, twenty geographically separate GRAND10K replicas will be built in hotspots for neutrino detection, with the sensitivity of 4×$10^{-10}$ GeV cm$^{-2}$ s$^{-1}$ sr$^{-1}$ \cite{ref5}. This article introduces the GP300 hardware design and the tests applied to it.

\section{GRANDProto300 hardware}
\subsection{Detection unit}

\begin{figure}[htbp]
\centering
\includegraphics[width=0.4\linewidth]{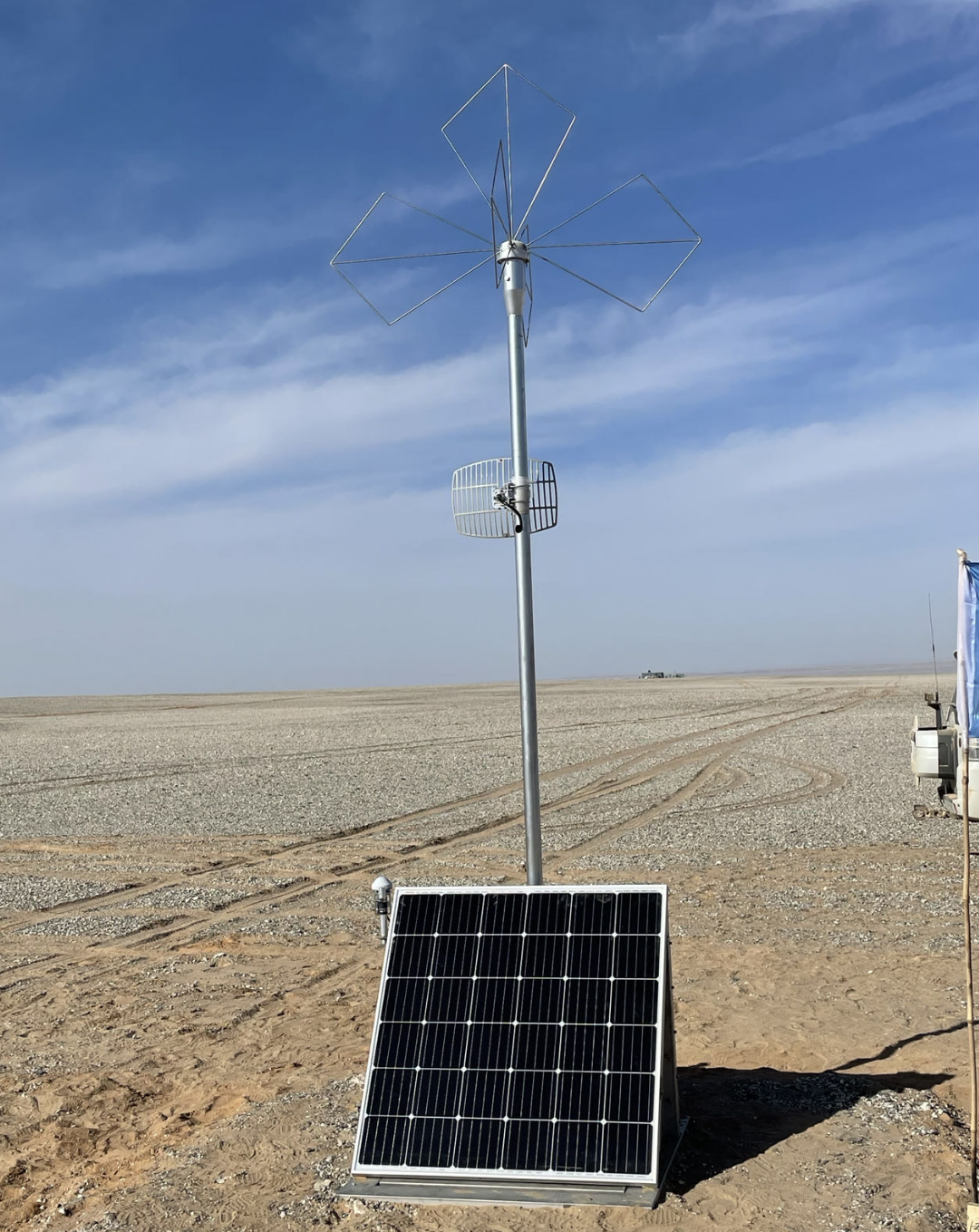}       
\caption{GRAND detection station.\label{Fig.2}}
\end{figure}

Figure \ref{Fig.2} shows a GRANDProto300 detection unit in the field. Each detection unit is autonomous: they generate their own power and communicate through wireless connections. The station consists of a receiving antenna installed on a 3.5-m pole. The current design is composed of three butterfly-shaped antennas in three polarization direction: two perpendicular dipoles in the horizontal plane and a monopole for the vertical polarization. A low-noise amplifier is mounted below the antenna and a wireless communication antenna is also installed halfway up the pole. The pole is supported by a box covered with a solar panel that provides power to the system. The battery is used internally to store and distribute electricity for electronic devices for digitization, analysis, triggering, and communication to the central data acquisition (DAQ) system. Below the triangular box holding the electronics is a rectangular one that can be filled with sand and rock on-site to mechanically stabilize the structure in strong winds \cite{ref3}.

\subsection{DAQ board}

\begin{figure}[htbp]
\centering
\includegraphics[width=0.7\linewidth]{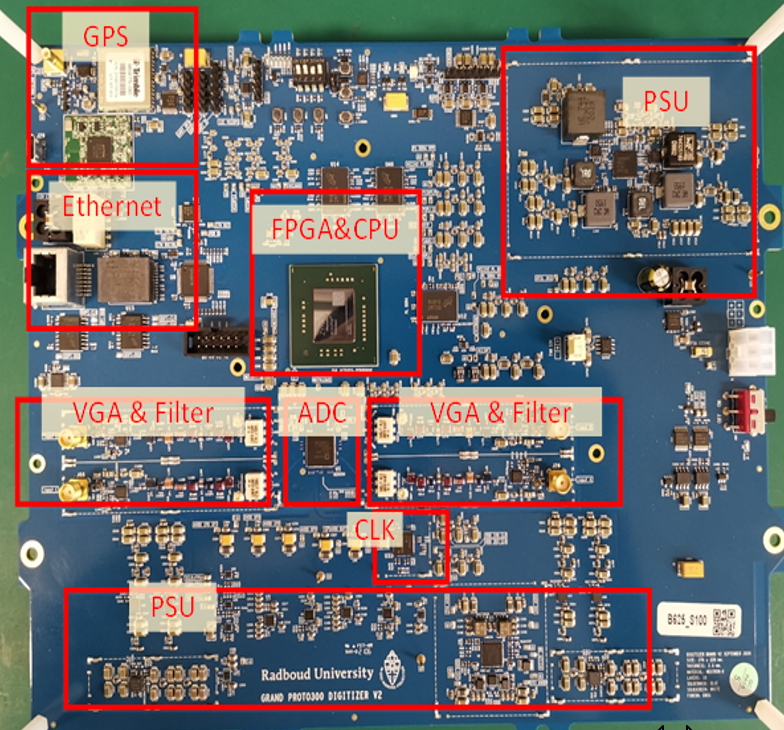}       
\caption{GRAND DAQ board.\label{Fig.3}}
\end{figure}

Figure \ref{Fig.3} shows the data acquisition (DAQ) board inside one of the detection units. There, a recorded signal is amplified, filtered, and digitized. The detection frequency range is 30--230 MHz, which is the passband range of the bandpass filter. An analog-to-digital conveter (ADC) AD9694 digitizes the signal on 14 bits at a rate of 500 MSamples/s. A Zynq UltraScale+ MPSoC that integrates FPGA and Arm CPU processes the signal, provides a trigger and communicates with a central DAQ station. More details can be found in \cite{ref6}.

\section{Tests of the front-end DAQ board}

\subsection{Test environment}

\begin{figure}[htbp]
\centering
\includegraphics[width=0.7\linewidth]{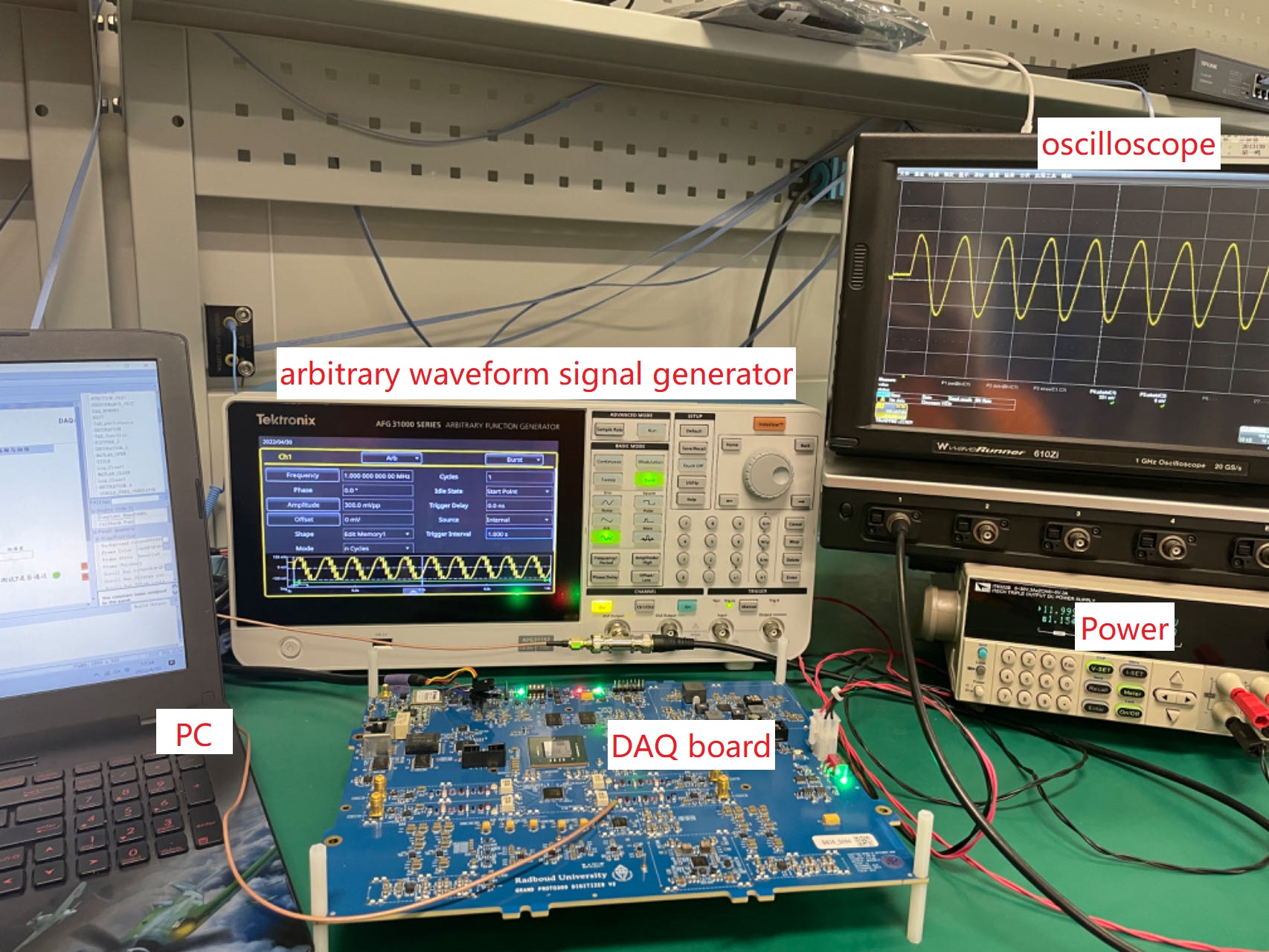}       
\caption{Test environment in laboratory.\label{Fig.4}}
\end{figure}

Figure \ref{Fig.4} builds the environment built to test the DAQ board in the laboratory. An arbitrary waveform generator, controlled by a computer, injects signals into the SubMiniature version A (SMA) interface of the DAQ board. The digitized signals are then displayed and analyzed in real-time on the DAQ computer.

\subsection{Test system graphical user interface}

\begin{figure}[htbp]
\centering
\includegraphics[width=0.9\linewidth]{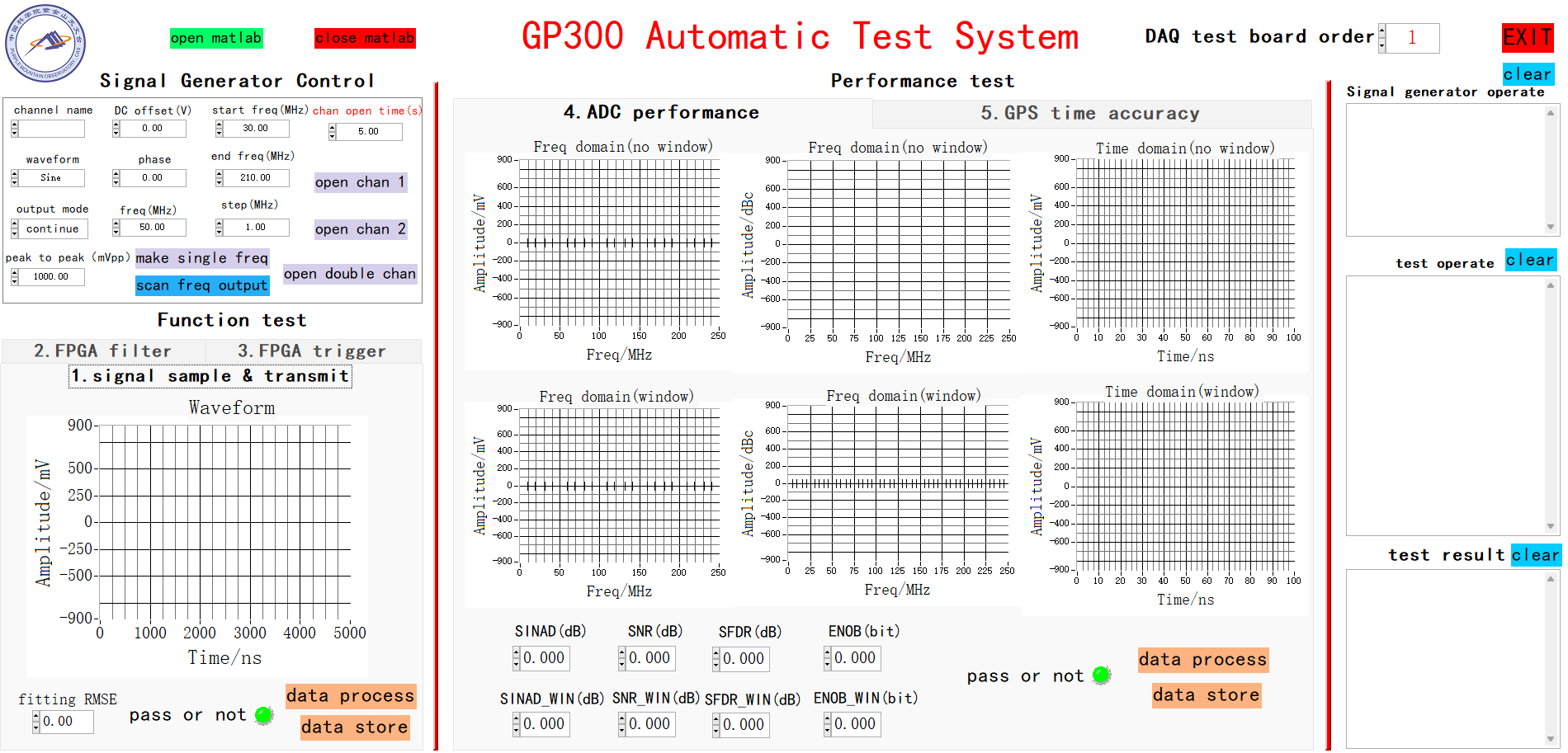}       
\caption{Test system graphical user interface.\label{Fig.5}}
\end{figure}

Figure \ref{Fig.5} shows the graphical user interface that we developed to test the board efficiently. The system is developed using LabWindows/CVI and is divided into three sections: signal generator, function test, and performance test.

\subsection{Functional test}

Function tests are of signal acquisition and transmission, and of the field-programmable gate array (FPGA) filter algorithm. 

\subsubsection{Signal acquisition and transmission}

\begin{figure}[htbp]
\centering
\includegraphics[width=0.7\linewidth]{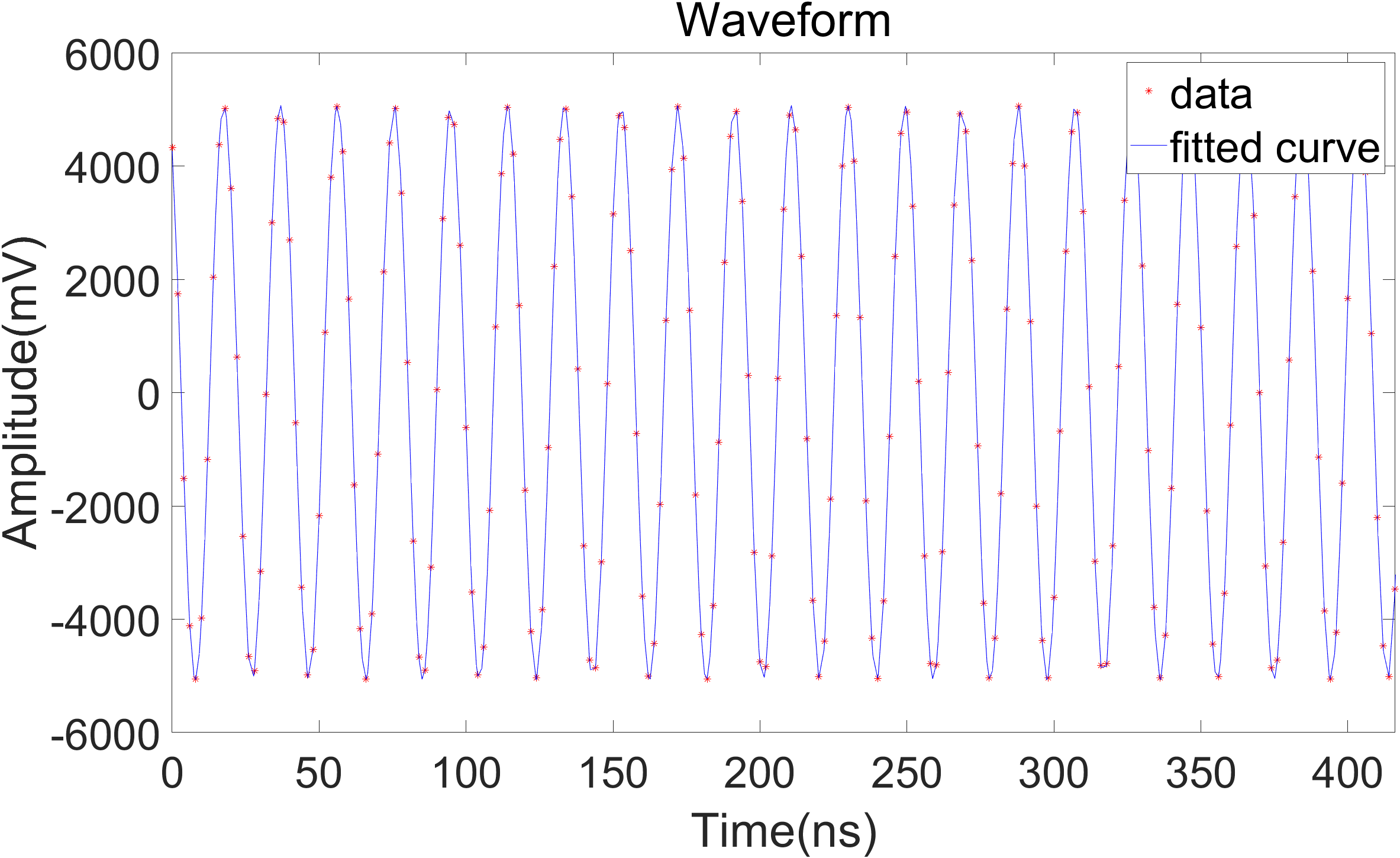}       
\caption{Digitized sine waveform injected into the DAQ board.\label{Fig.6}}
\end{figure}

The waveform generator injects a 51.78~MHz sine wave into the DAQ board, and the collected data is fitted as a sine wave using MATLAB; see Fig. \ref{Fig.6}.

\subsubsection{FPGA filter algorithm}

\begin{figure}[htbp]
    \begin{minipage}[t]{0.5\linewidth}
        \centering
        \includegraphics[width=\textwidth]{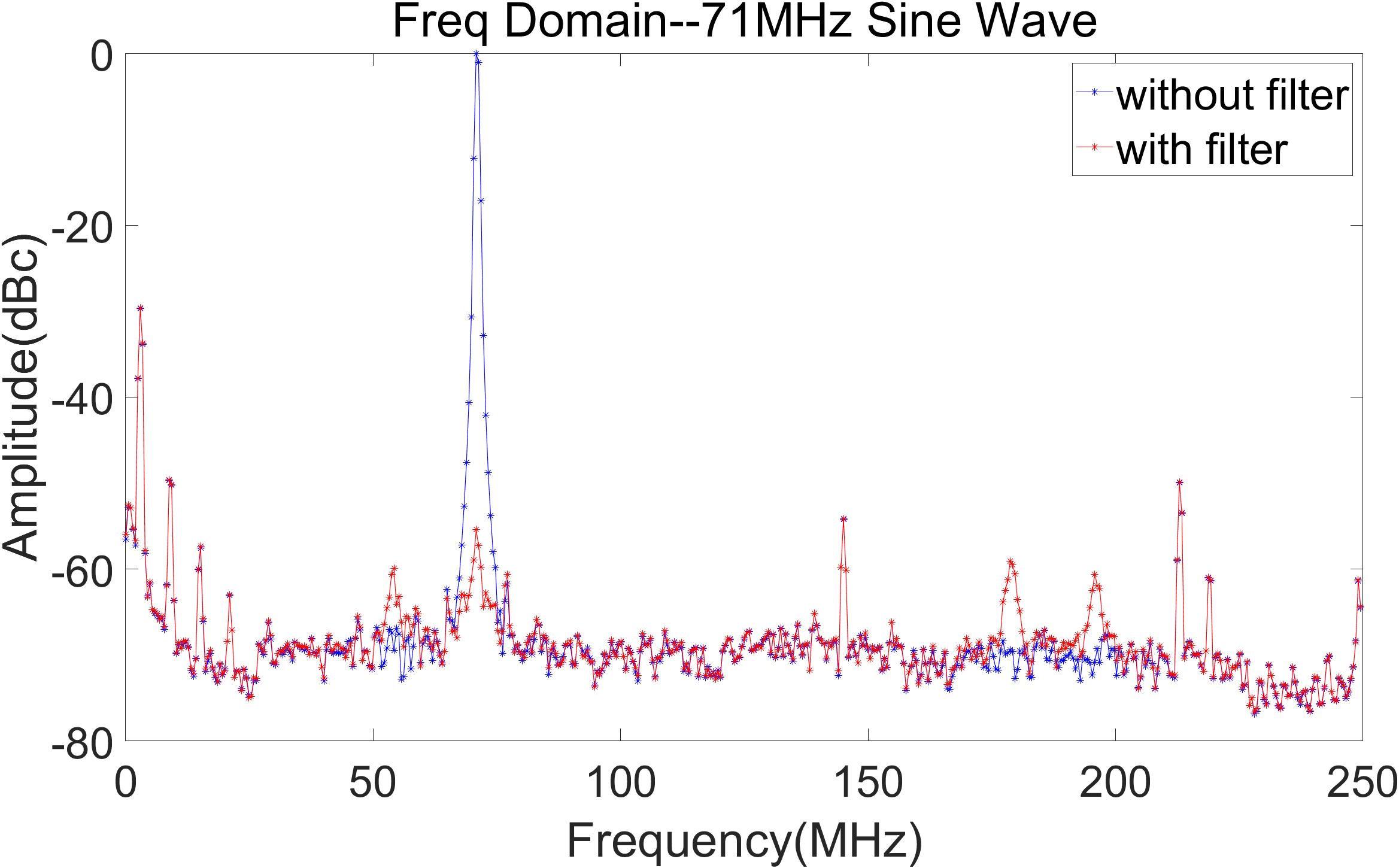}
        \centerline{(a) Input: 71~MHz sine wave}
    \end{minipage}
    \begin{minipage}[t]{0.5\linewidth}
        \centering
        \includegraphics[width=\textwidth]{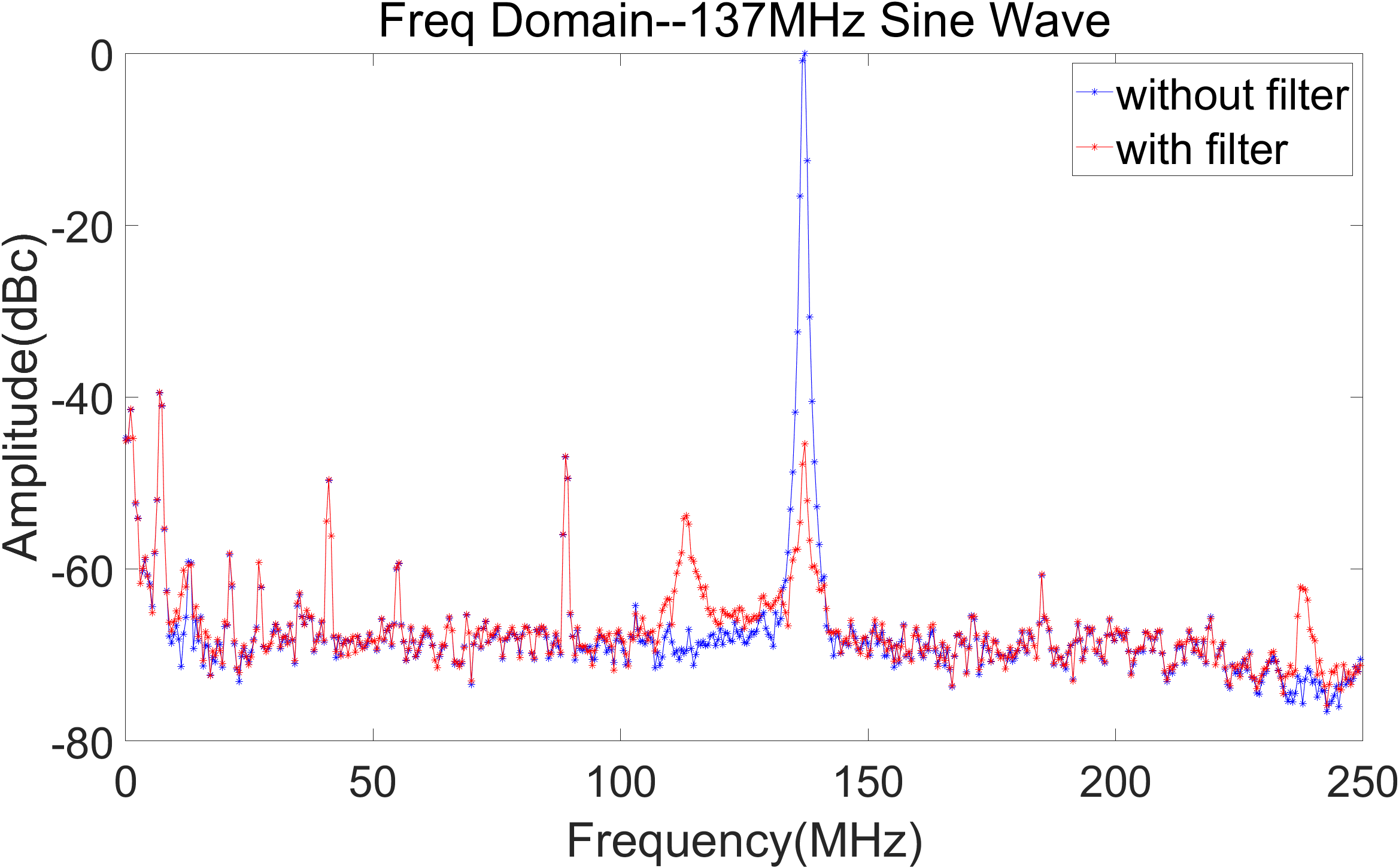}
        \centerline{(b) Input: 137~MHz sine wave}
    \end{minipage}
    \caption{The power spectra of a 71~MHz and a 137~MHz sine wave injected into the DAQ board.\label{Fig.7}}
\end{figure}

A sine wave is injected into the DAQ board using the waveform generator, and the collected data is shown in the frequency domain using a fast Fourier transform; see Fig. \ref{Fig.7}. When the frequency of the injected signal is 71~MHz, the center frequency of infinite impulse response notch filter is configured as 71~MHz and the filter depth is configured as 0.99 in FPGA, and the same applies to 137 MHz. The blue line represents the spectrum without the filter, and the red line represents the spectrum with the filter. The filtered signal strength is reduced by at least 40~dB.

\subsection{Performance test}

Performance tests are of the analog-to-digital conversion and the GPS time accuracy. Only one representative board was tested.

\subsubsection{ADC performance}

We used a radiofrequency and microwave signal generator (SMA100B) to inject a 115-MHz sine wave into the DAQ board. After data acquisition, the signal is analyzed in the frequency domain and the effective number of bits (ENOB) of the ADC can be calculated using the following formula \cite{ref7}: 
\begin{equation}
{\rm SINAD} = 20 \log_{10}{\frac{A_{\rm signal}}{A_{\rm noise}+A_{\rm distor}}}
\end{equation}
\begin{equation}
{\rm ENOB} = \frac{{\rm SINAD}-1.76~{\rm dB}}{6.02~{\rm dB}/{\rm bit}} 
\end{equation}
where SINAD is the signal-to-noise-and-distortion ratio, $A_{\rm signal}$ is the amplitude of the inejcted sine wave, $A_{\rm noise}$ is the amplitude of the noise, and $A_{\rm distor}$ is the amplitude of the distortion. The ENOB of the ADC is 10.69, which is consistent with the range of ENOB measured in the data sheet of the ADC \cite{ref8}.


\subsubsection{GPS time accuracy}
Synchronization signals are injected into two DAQ boards from an arbitrary waveform generator. The timing difference between the two boards for the same event are counted; see Fig.~9. The trigger interval is 1~s, and 2000 events are acquired. The test duration is approximately 33 minutes. The time difference is fitted by a Gaussian, with a standard deviation of 11.62~ns. The mean value of the distribution is 71~ns and fluctuates up and down by 90~ns. When I change the test duration, the mean value will vary. The specific reason needs further discussion.

\begin{figure}[htbp]
    \begin{minipage}[t]{0.5\linewidth}
        \centering
        \includegraphics[width=\textwidth]{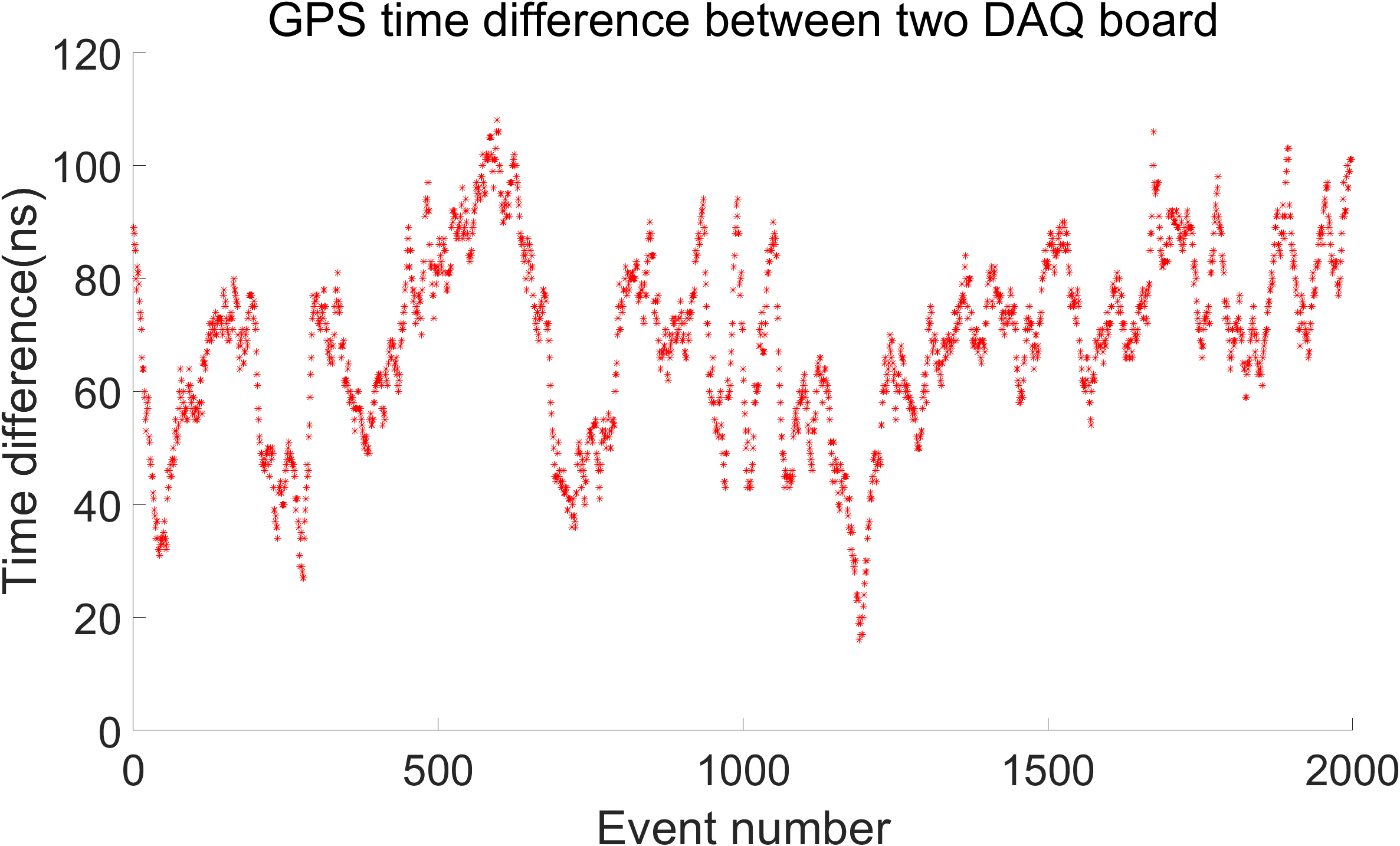}
        \centerline{(a) the time difference between two DAQ boards}
    \end{minipage}%
    \begin{minipage}[t]{0.5\linewidth}
        \centering
        \includegraphics[width=\textwidth]{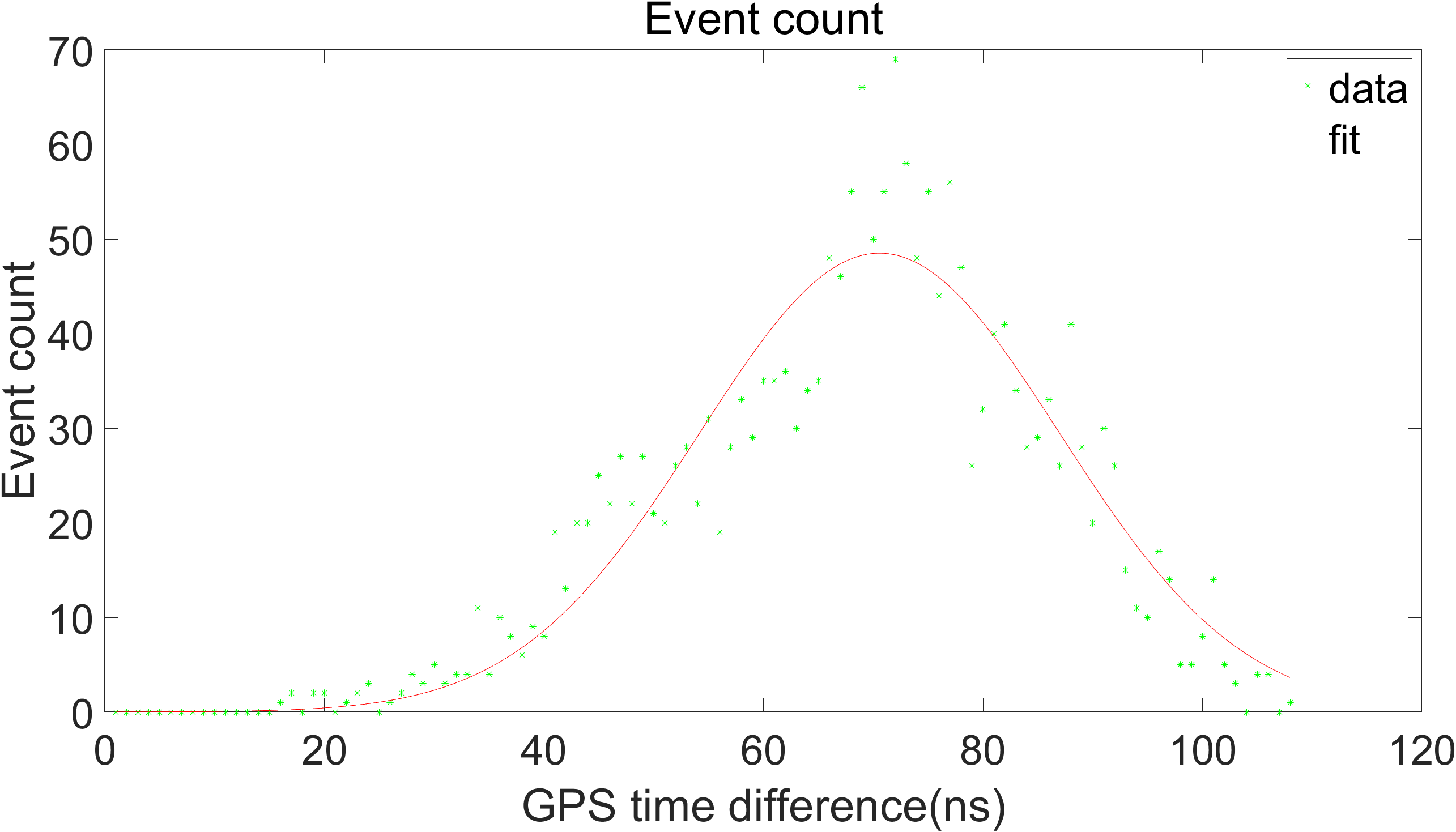}
        \centerline{(b) Event count}
    \end{minipage}
    \caption{GPS time difference between two DAQ boards\label{Fig.9}}
\end{figure}

\section{Conclusion}
GRANDProto300 is the 300-antenna array that serves as a pathfinder to GRAND. We have developed a system to test the functionality and performance of the data acquisition boards housed by the GRANDProto300 detection units. Our tests find that the boards live up to the requirements of GRAND.

%
%
%

\end{document}